\documentclass[a4paper,10pt,twocolumn,twoside,aps,prb,superscriptaddress]{revtex4-2}
\usepackage{multirow}
\usepackage[utf8]{inputenc}
\usepackage{graphicx}
\usepackage{amsmath}
\usepackage[svgnames]{xcolor}
\graphicspath{{./figuras/}}
\usepackage[caption=false,subrefformat=parens,listofformat=parens]{subfig}
\usepackage{hyperref}
\hypersetup{
     colorlinks = {true},
     linkcolor = {blue},
     anchorcolor = {blue},
     citecolor = {blue},
     filecolor = {blue},
     urlcolor = {blue},
     pdfauthor = {C. Helman},
     pdftitle = {Influence of Fermi surface shape on magnetotransport },
     }

\begin{document}
\title{Influence of the Fermi surface shape on magnetotransport: the MnAs case}
\date{\today}

\author{C. Helman}
\affiliation{Centro Atómico Bariloche, Comisión Nacional de Energía Atómica, S. C. de Bariloche, CP8400, Pcia de Rio Negro, Argentina}
\email{christianhelman@cnea.gov.ar}
\author{A.M. Llois}
\author{ M. Tortarolo}
\affiliation{Instituto de Nanociencia y Nanotecnología, CONICET-CNEA. Centro Atómico Constituyentes, Av. Gral. Paz 1499, B1650 Villa Maipú, Buenos Aires, Argentina}

\begin{abstract}


We analyze the influence of the Fermi surface (FS) shape on magnetotransport properties, particularly on the Hall effect in the MnAs compound. 
It has been observed in MnAs films evidence of opposite conduction polarities  for different crystal direction (\textit{goniopolarity}) and a strong dependence of the carrier type with applied magnetic field.
In order to understand this behaviour, we developed a model based on the semiclassical equations along with Boltzmann transport theory that takes into account  both, the applied magnetic field and the FS shape. 
The FS of the MnAs compound is obtained by means of density functional theory (DFT), showing a clear dominance of the hyperboloid shape. 
Our study, corroborate that this specific topology of the FS gives rise to a \textit{goniopolar} behaviour in the Hall transport.
This theoretical results are supported by magnetotransport measurements on MnAs thin layers epitaxially grown on GaAs(001) and GaAs(111), where both configurations  allow us to explore the transport characteristics for two different crystal directions of the MnAs. 

\end{abstract}
\maketitle

\section{Introduction}
The shape of the Fermi surface (FS) has been used to describe transport properties since the early days of material science.
Pioneering works by Pippard \cite{Pippard1960} and Lifshitz \cite{1955} theoretically prove that the magnetotransport behavior depends on the type of orbits that the wave vector performs in momentum space.
These orbits are classified as open or closed only between two scattering events, with $\tau$ being its characteristic time.
For these cases, the magnetoresistance(MR) and especially the Hall resistivity present different behavior with the magnetic field.
On one hand, closed orbits lead to a linear dependence of the Hall resistivity and a saturating MR with increasing applied magnetic field. If these orbits on the FS enclose a region with lower energy, then positive values are expected for Hall resistivity (hole-type) and negative values on the opposite case (electron-type).
On the other hand, open orbits lead to a quadratic dependence  of the Hall resistivity (hole type) and non-saturating magnetoresitance with the applied magnetic field.

The above-mentioned behavior of closed/open orbits in momentum space are usually recovered from the FS modeled as an ellipsoid connected with a neck that reaches the boundary of the Brillouin zone (BZ) \cite{Gurzhi1975, Gurzhi1977, Pippard1960}.
Other attempts to describe the magnetotransport properties use a tubular-like FS model, but the analysis emphasizes on the connectivity of the surface more than the concavity \cite{Pippard1960, Novikov2019}.
Recently, \textcite{He2019} shows that the concavity of the FS leads to simultaneous electron and hole-like magnetotransport properties, especially when presenting open orbits.
Hence, for different crystal directions, the transport properties show opposite conduction polarities.
This effect, denominated \textit{goniopolar} by the authors, is theoretically and experimentally studied in NiSn$_2$As$_2$ \cite{He2019,Wang2020,Lambrecht2020}.
The origin of this phenomenon is straightforward in the case of a single band FS when it is singly connected with open orbits in one crystal direction and  closed orbits in other direction.
The different kind of orbits give rise to a noticeable different behaviour in electron transport measurements for different crystal directions \cite{Zhang2019}.
In the case of the Hall effect (HE), the \textit{goniopolarity} is manifested when the Ordinary Hall coefficient R$_H$ has the opposite sign when measuring it with respect to different crystal orientations as a function of the applied magnetic field.
Similar behavior is also observed in the case of the Seebeck effect when the thermopower has the opposite sign for different crystalline orientations\cite{He2019}.

Manganese Arsenide (MnAs) compound is a good candidate to be classified as a \textit{goniopolar} material because it presents simultaneously electron- and hole-like transport behavior, as earlier observed by \textcite{Berry2001}.
Their MR and the HE measurements in MnAs/GaAs(001) epilayers reveal the presence of both electron and holes in the magnetotransport, with a contribution that varies with temperature and magnetic field.
Also, \textcite{Friedland2003} observed mainly the same characteristics in the MnAs/GaAs
magnetotransport, adding that the carrier type strongly depends on the crystal orientation: MnAs/GaAs(001) samples
exhibit mixed hole like and electron like conductivity already at zero magnetic field, while in MnAs/GaAs(111) the low-temperature transport is dominated by holes at zero magnetic fields.

To explain their results Ref. \cite{Berry2001} propose a two carrier model, while Ref. \cite{Friedland2003} use a model based on spherical bands with a small number of impurities \cite{Berger1969}.
Nowadays, we know that MnAs compound shows complex Fermi surfaces topologically different from spherical models, and the good quality of the MnAs/GaAs epilayers \cite{Daweritz2001,Ramsteiner2002,Mattoso2004} diminishes the role of impurities and domains in low-temperature transport phenomena.

In this work, we analyze the magnetotransport properties of MnAs from both experimental and theoretical points of view.
In Sec. II we present the necessary theoretical background to explain our transport measurements.
Then Sec. III A describes the MnAs compound using the \textit{ab initio} calculations and presents the FSs from which we modeled the magnetotransport behavior, and Sec. III B presents our measurements (MR, HE and magnetization) on MnAs/GaAs in different crystal orientations.  Sec. IV is a discussion on the theoretical model presented and its accuracy to describe the experimental measurements.

\section{Theoretical background\label{sec:teoria}}
In order to describe the transport behavior with magnetic field we restrict ourselves to a semiclassical treatment, in which the electrons can be thought as classical particles obeying Fermi–Dirac statistics.
At low temperature the features of the FS rule the electronic properties of metals, where the mean free time $\tau$ can be assumed to be large and only band dependent.
This extreme case in temperature can be achieved in our theoretical approach by tending the smearing of the FS occupation function to zero in the Boltzmann transport equation.
Under these conditions the solution for the conductivity within the relaxation time approximation (RTA) is \cite{Ashcroft1976}
\begin{equation}
\label{eq:conductivity}
 \sigma_{i,j}=e^2 \sum_n \tau^{(n)} \int_{FS^{(n)}} v_i^{(n)}(\mathbf{k})  \bar{v}_j^{(n)}(\mathbf{k})\frac{d\mathbf{S}}{|\nabla\varepsilon^{(n)}(\mathbf{k})|},
\end{equation}
where i,j are the cartesian coordinates referred to the crystal axis, $e$ is the electron charge and $\tau^{(n)}$ is the band dependent relaxation time.
The index $n$ indicates the band number, while the sum is over the bands that cross the Fermi level.
The factor $1/|\nabla\varepsilon^{(n)}(\mathbf{k})|$ is related to the density of states of the $n$-th band with energy $E_{F}$.
The integral over the FS of the $n$-th band ($FS^{(n)}$) is parameterised as $\varepsilon^{(n)}(\mathbf{k})=E_F$ .

The integrand of the  Eq.~(\ref{eq:conductivity}) involves a product of two different kind of velocities:
$ \mathbf{v}^{(n)}(\mathbf{k})$ is the  gradient in momentum space of the energy band, and $\bar{\mathbf{v}}^{(n)}$ is defined as the weighted average over the past history of the charge carrier,
\begin{equation}
\label{eq:promedio}
 \bar{v}_{j}^{(n)}=\int_{-\infty}^0\frac{e^{t/\tau^{(n)}}}{\tau^{(n)}} v_{j}^{(n)}(\mathbf{k}(t)) dt.
\end{equation}

$\mathbf{k}(t)$ is the group velocity and its time evolution in momentum space due to applied magnetic field, ($\mathbf{H}$), is derived from the semiclassical set of  equations (assuming no band-crossing),
\begin{subequations}
\label{eq:motion}
\begin{eqnarray}
\mathbf{v}^{(n)}=&\frac{1}{\hbar}\nabla_{\mathbf{k}}\varepsilon^{(n)}(\mathbf{k}) , \label{eq:motion1}\\
 \hbar\dot{\mathbf{k}}(t)=&-\frac{e}{c}\mathbf{v}^{(n)}(\mathbf{k}(t))\times\mathbf{H},\label{eq:motion2}\\
 \mathbf{k}(t=0)=&\mathbf{k}_0 \in FS^{(n)}\label{eq:motion3}.
 \end{eqnarray}
\end{subequations}

Note that the set of Eqs.\eqref{eq:motion} must be solved for each band that crosses the Fermi level and its solution has to be inserted in Eq.\eqref{eq:promedio}.
While $\mathbf{v}^{(n)}$ only depends on the wave vector $\mathbf{k}$, $\bar{\mathbf{v}}^{(n)}$ depends on both $\mathbf{k}$ and the magnetic field $\mathbf{H}$.

Two conservation laws can be derived from the set of Eqs.\eqref{eq:motion}; $i$) the wave vector trajectory in the momentum-space is the intersection of the FS with a plane perpendicular to the magnetic field direction (Fig.1); and $ii$) the energy remains constant in the presence of an  applied magnetic field.
Both conservation laws allow us to have a preliminary picture of the results for different FSs.

The resistivity tensor is obtained from the relation $\mathbf{\rho}=\mathbf{\sigma}^{-1}$, which in principle requires a detailed description  of $\sigma$.
However,  if we take into account the crystal symmetries of the compound and the Onsager's relation ($\rho_{i,j}(\mathbf{H})=\rho_{j,i}(-\mathbf{H}) $), then an interrelation emerges among tensor components that reduces the number of calculation required to obtain $\sigma$  \cite{Akgoz1975}.

\subsection*{Hyperboloid Fermi surface}
Many semiclassical transport calculations have been done taking into account spherical-like FS, with or without open orbits\cite{Pippard1960,1955,Ashcroft1976,Gurzhi1975,Gurzhi1977,Zhang2019}.
In this section we analyse the case of a FS that is concave in one direction and convex in other one. More precisely, a circular hyperboloid of one sheet or an hyperboloid of revolution, which leads to a noticeably different behaviour in conductance (or resistivity), as we discuss next.

\begin{figure}[b]
\centering
\includegraphics[scale=0.4]{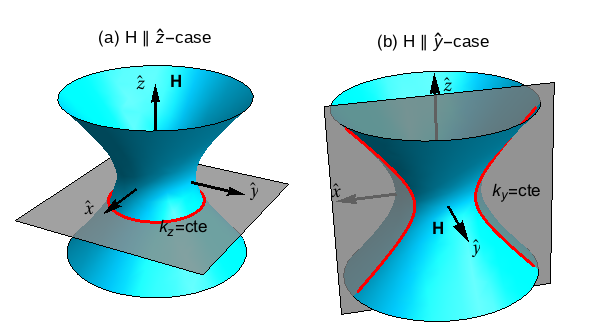}
 \caption{\label{fig:hyper}Hyperboloid of revolution as a FS model from Eq. \protect\ref{eq:fs}. The applied magnetic field is perpendicular to the gray planes  and the intersection with FS is in red. (a) $\mathbf{H}\parallel\hat{z}$, only close orbits are possible.  (b) $\mathbf{H}\parallel\hat{z}$, where  open orbits are allowed. }
\end{figure}

Assuming a FS parameterised as
\begin{equation}
\label{eq:fs}
 \varepsilon(\mathbf{k})=\frac{\hbar^2}{2}(\frac{k^2_x}{m_x}+\frac{k^2_y}{m_y}-\frac{k^2_z}{m_z})=E_F,
\end{equation}
where $m_i$ are the effective masses. The minus sign on the $k^2_z$-term indicates that the rotation symmetry axis is parallel to $z$-axis, while $m_x=m_y$ is the condition to have a hyperboloid of revolution.
For our further analysis it is convenient to define the quantity $\alpha=\frac{m_z}{m_x}$ that is always a real positive number.
If $\alpha\gg1$, the hyperboloid surface approaches to a cylinder with small concave curvature, while for $\alpha\approx0$, the surface has two parabolic sheets joined by a narrow neck.
Also, it is well known that bands must be perpendicular to the BZ boundaries.
To fulfill this requirement, we can define energy as a continuous and piece-wise function.
However, the change in the energy function from convex to concave is very small and its effect on the orbits is negligible \cite{He2019}.

In order to obtain $\bar{\mathbf{v}}^{(n)}$,  we need the time-dependent wave-vector expression  $\mathbf{k}(t)$, which describes the orbits on the FS.
Since $z$-axis is a center of rotational symmetry, only two cases are important to analyse: One where the magnetic field $\mathbf{H} \parallel$z-axis and the other one when $\mathbf{H}$ lies in the hexagonal planes, \textit{i.e}. $\mathbf{H}\parallel$y-axis.
The aforementioned conservation law can help us to have an insight on the solution of the set of Eq. \eqref{eq:motion}.
For the $\textit{H$\parallel$z}$ case, the obtained orbits are closed circles, as shown in Fig. \ref{fig:hyper}(a), where the time dependent solution has the characteristic frequency $\omega$$_{z}$=$\frac{e}{c}\frac{H}{m_x}$.
Instead, for the H$\parallel$y case, the orbits are open hyperboloid, as shown in Fig. \ref{fig:hyper}(b), with characteristic frequency $\omega_{y}=\frac{e}{c}\frac{H}{m_x\sqrt{\alpha}}=\omega_z/\sqrt{\alpha}$.
For simplicity we use $\omega_{z,y}$ instead magnetic field $\mathbf{H}$.
In the case of closed orbits, the physical meaning of $\omega$ is the inverse of the time to do one cycle. Hence, for higher magnetic field the wave-vector can perform more cycles.
A similar concept can be induced in the case of open orbits, a larger length of the orbit in the extended BZ is covered with an increasing magnetic field.
In our approach, the RTA and low field condition happen when $\omega\tau<1$, which is the main difference with previous works of Ref. \cite{Pippard1960,1955} and more recently \cite{Novikov2019}, where they analyse the high field limit ($\omega\tau\gg1$).

The expression of $\bar{\mathbf{v}}^{(n)}$ depends on $\omega\tau$, the effective mass $m_x$, the mass parameter $\alpha$, and the initial position of the wave-vector $\mathbf{k}_0$, which could be any point on the FS as pointed in  Eq. \eqref{eq:motion3}.
In order to obtain the final expression of the conductivity we must integrate over the FS. In this process the vector $\mathbf{k}_0$ becomes an integration variable and the final expression takes into account the history of all possible orbits on the FS.

The resistivity tensor for the H$\parallel$y-configuration can be expressed as a function of  $\alpha$ and $\omega_z\tau$ (for simplicity we omit the index $z$ ),
\begin{equation}\label{eq:abiertas}
 \rho^{open}=\rho_0
\begin{pmatrix}
 \frac{\alpha-(\omega\tau)^2}{g_1+(\omega\tau)^2} &0 & \omega\tau \frac{\alpha-(\omega\tau)^2}{g_2+(\omega\tau)^2} \\
 0&1&0\\
 -\omega\tau \frac{\alpha-(\omega\tau)^2}{g_2+(\omega\tau)^2}&0&\alpha\frac{\alpha-(\omega\tau)^2}{g_3+(\omega\tau)^2}
\end{pmatrix}.
\end{equation}
In this configuration the current is perpendicular to the applied magnetic field and it can be in $\hat{x}$ or $\hat{z}$ directions.
Instead, for the H$\parallel$z-case, the resistivity tensor is the expected for closed orbits;
\begin{equation}
 \rho^{closed}={\rho'}_0
\begin{pmatrix}\label{eq:cerradas}
 1 & \omega\tau &0 \\
 -\omega\tau & 1 &0\\
 0&0&g_4
\end{pmatrix}.
\end{equation}
Both tensors are antisymmetric and follow the Onsager's relations.
We group most of the constant in $\rho_0$, while $g_i$ are geometric factors that come from FS integrals.
The off diagonal elements of $\rho^{open}$ and $\rho^{closed}$ are the so called ordinary Hall resistivity, which is obtained in the frame of semiclassical theory.

\begin{figure}[t]
 \includegraphics[scale=0.5]{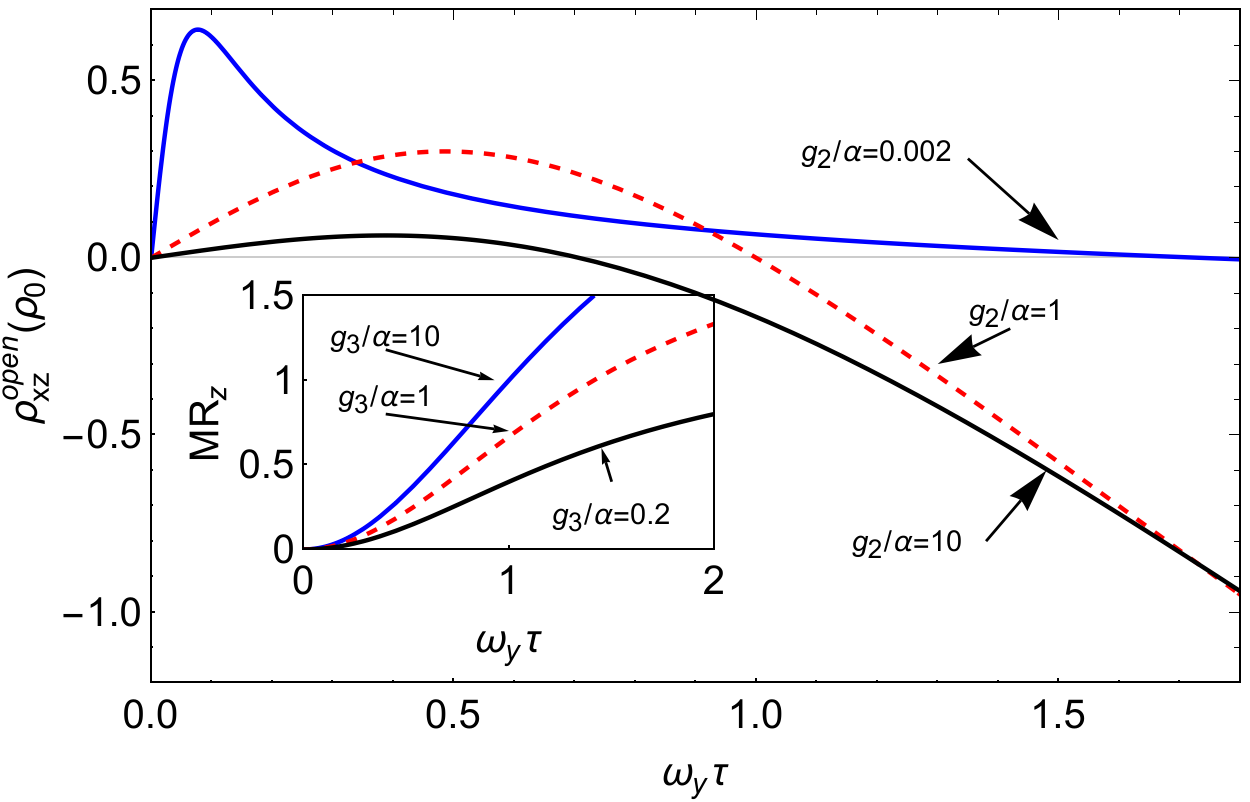}
\caption{\label{fig:rhoxy}
Curves corresponding to Eq.\eqref{eq:abiertas} as a function of $\Omega\tau$, where  H$\parallel y$-axis (see Fig.\ref{fig:hyper}).
Ordinary Hall is in units of $\rho_0$ and magnetoresistance (inset) assume current in $z$-direction.
The different curves correspond to values of $g_2/\alpha$ with $\alpha=\frac{m_z}{m_x}$ and $g_2$ is a geometric factor that comes from integral of Eq.\eqref{eq:conductivity}. }
\end{figure}

For the case H$\parallel$y, we plot in Fig. (\ref{fig:rhoxy}) the ordinary Hall resistivity $\rho^{open}_{xz}$ and the MR (inset) for different values of the parameter $g2/\alpha$.
The ordinary Hall has a root when $\omega_y\tau=1$ or equivalently $\omega_z\tau=\sqrt{\alpha}$, and then it changes the sign, which can be interpreted as a switch of the carrier type, from holes to electron.
In all cases of $g_2/\alpha $, the ordinary Hall $\rho^{open}_{xz}$, presents a mixed behavior with applied magnetic field.
For values of $\omega\tau<1$, the Hall resistivity is positive and reaches a maximum value that depends on $\alpha$.
If $g_2/\alpha=1$, then the maximum of  $\rho_{Hall}$ is reached for $\omega\tau\approx0.5$, higher values of $g_2/\alpha$ displace the maximum to zero, and for smaller values the maximum approaches 1.
The MR defined as $|\rho_{ii}(\mathbf{H})-\rho_{ii}(0))|/|\rho_{ii}(0)|$, saturates for values of $\omega\tau$ where the RTA and low field condition are no longer valid ($\omega\tau\gg1$).
In all the cases of $g_3/\alpha$, the MR presents a quadratic-like behaviour for $\omega\tau<1$, as expected for open orbits.

The \textit{goniopolar} behavior is manifested by comparing Eq. \eqref{eq:abiertas} and \eqref{eq:cerradas}, where a different magnetotransport behaviour should be expected for this kind of FS.
When the magnetic field is parallel to the $y$-axis (H$\parallel$y), the ordinary Hall resistivity as a function of the applied magnetic field presents a non-monotonic behavior, having positive slope for lower $\omega_z\tau$ values and negative slope for higher ones.
Instead, when the magnetic field is parallel the $z$-axis (H$\parallel$z), the ordinary Hall resistivity presents a linear behavior with increasing field.

Nevertheless, the present analysis is for a single band and simply connected FS sheet.
In the case of multiband transport, the effects that come from other FS sheets should be taken into account in order to reproduce the experimental results.
However, the curves obtained for the hyperboloid model, specially for $g_2/\alpha\approx1$, present the same behaviour with applied magnetic fields as our experimental measurements \ref{sec:exp} on MnAs/GaAs epilayers, and similar ones that were reported \cite{Berry2001, Friedland2003}.
These results lead us to study the FS of the MnAs compound.

\section{The MnAs compound}
\begin{table}[t]
 \caption{\label{tab:estructura}Relation among components of the resistivity for the $P63/mmc$ space structure. The figure shows the atomic arrangements as well as the lattice parameters.}
 \begin{tabular}{c c l}
  Structure & Direction & Relation\\
  \hline
  \multirow{8}{0.16\textwidth}{\includegraphics[scale=0.3]{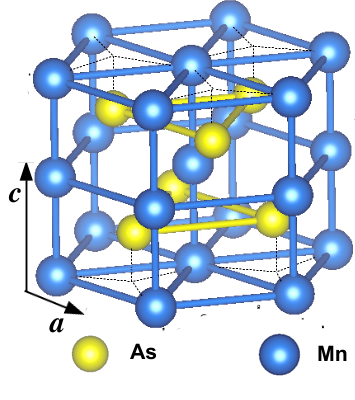} } & & \\
   &$\mathbf{H}\parallel c$-axis & $\rho_{xz}=\rho_{yz}=0$ \\
   & (H$\parallel$z) & $\rho_{xy}(H)=-\rho_{yx}(-H)$ \\
   \\
   & $\mathbf{H}\parallel ab$-plane &  $\rho_{xy}=0;\rho_{xx}=\rho_{yy}$\\
   & (H$\parallel$y) & $\rho_{xz}(H)=-\rho_{zx}(-H)$     \\
   \\
   \\
 \end{tabular}
\end{table}
MnAs is ferromagnetic at room temperature and can be grown by Molecular Beam Epitaxy (MBE) onto several technologically relevant semiconductors as GaAs and Si ~\cite{Tanaka2002,Tanaka1994,Mattoso2004}.
Its ferromagnetic $\alpha$ phase crystallizes in the hexagonal NiAs structure with space group $P63/mmc$  for temperatures lower than 300K as shown in Table \ref{tab:estructura} \cite{Rungger2006}.
The magnetic easy axis lies in the hexagonal plane $ab$ and the hard direction is parallel to the $c$-axis.
Table \ref{tab:estructura} shows a scheme of the structure as well as the relations between components of the resistivity tensor due to the symmetries of the crystal structure.

\subsection{Electronic properties\label{sec:electronic}}
Following the early suggestion in Ref. \cite{Friedland2003}, that the features of the FS are responsible of the Hall resistivity behaviour with applied magnetic field, we obtain the FS by \textit{ab initio} band-structure calculations by means of the Quantum Espresso code (QE)\cite{Giannozzi2017}.
We use the well known generalized gradient approximation (GGA-PBE) for the exchange correlation potential with 8000 points in reciprocal space at the first Brillouin zone.
Since we want to describe the system for magnetic fields where the magnetization is saturated in a specific crystal direction, our calculations take into account the spin-orbit coupling (SOC) as implemented in the QE code and we use the fully relativistic pseudopotential for both, Mn and As atoms \cite{*[{Convergence achieved with 50 Ry for Energy cutoff of wave-function }]  [{Pseudopotentials used are the ones from the QE web-page.}] Giannozzi2017}.

Our \textit{ab initio} calculations are restricted to the two alignements consistent with the experimental geometries determined by the substrate orientation (Sec. \ref{sec:exp}).
Table \ref{tab:estructura} shows both magnetic moment configuration, that can be aligned with the $c$-axis of the crystal structure (H$\parallel$z-case), or to the $b$-axis (H$\parallel$y-case).

\begin{figure*}
\begin{minipage}{0.45\textwidth}
 \subfloat[ \label{fig:fermi18}]{
   \includegraphics[scale=0.3]{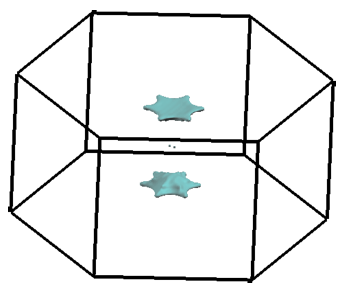}
   }
     \subfloat[\label{fig:fermi21}]{
    \includegraphics[scale=0.3]{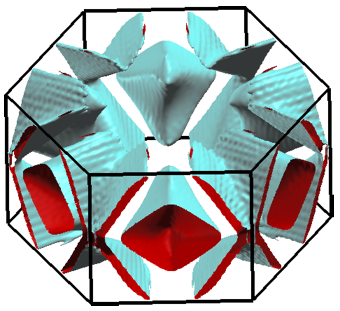}
    }\\
\subfloat[\label{fig:fermi20}]{
   \includegraphics[scale=0.3]{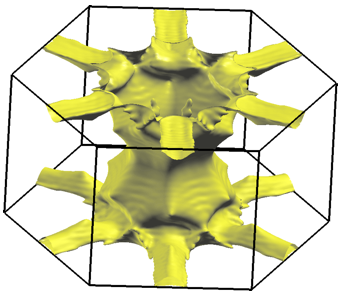}
    }
\subfloat[ \label{fig:fermi19}]
   {   \includegraphics[scale=0.28]{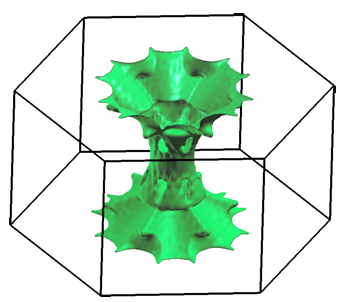}
    }
\end{minipage}
\begin{minipage}{0.45\textwidth}
 \subfloat[ \label{fig:ext1}]{
   \includegraphics[scale=0.85]{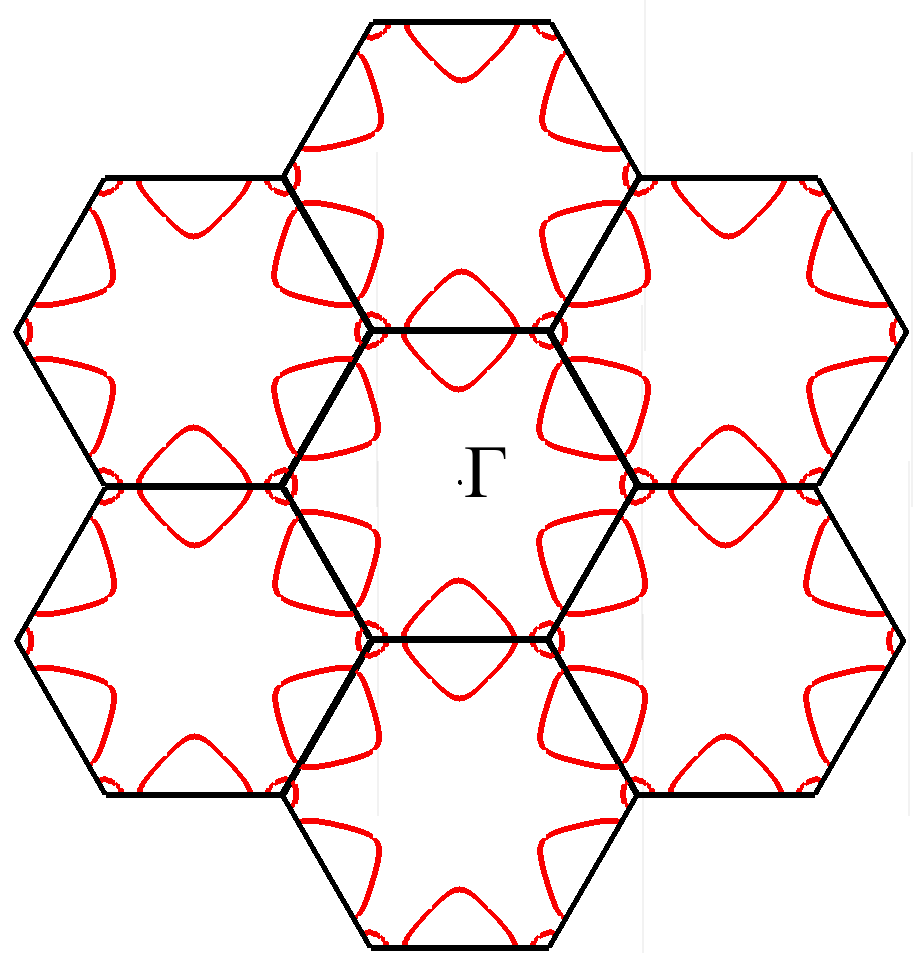}
   }
     \subfloat[\label{fig:ext2}]{
    \includegraphics[scale=1]{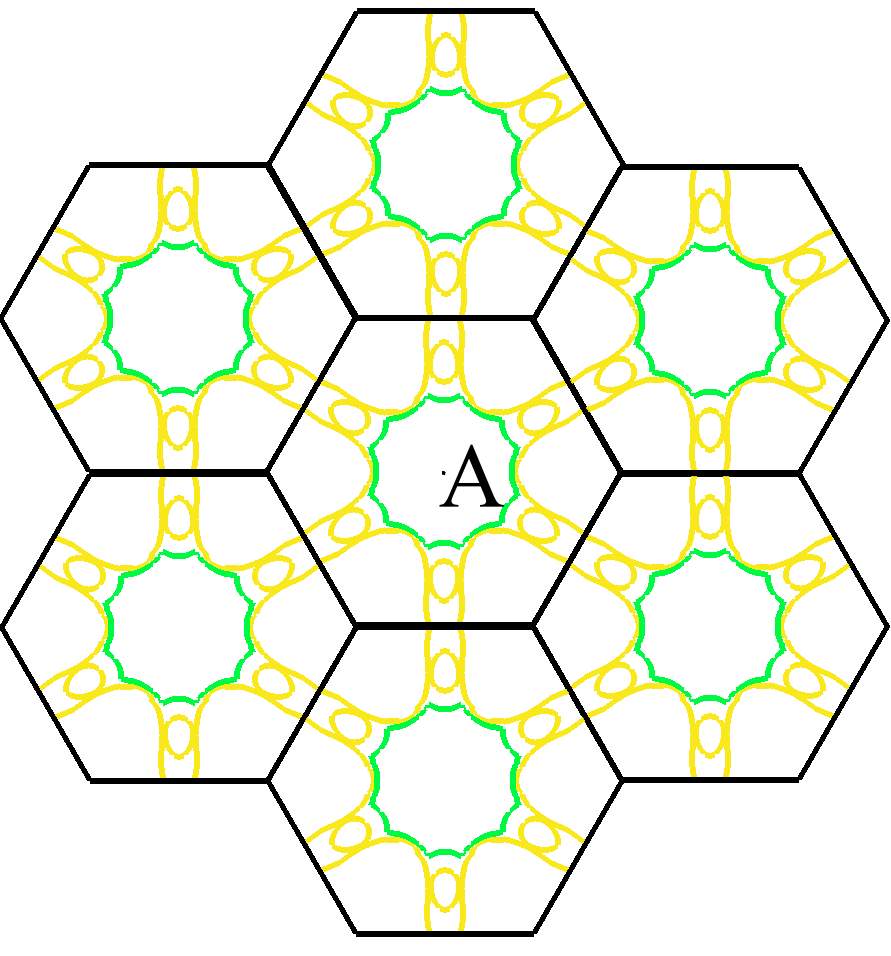}
    }\\
\subfloat[ \label{fig:proj1}]
   {   \includegraphics[scale=1]{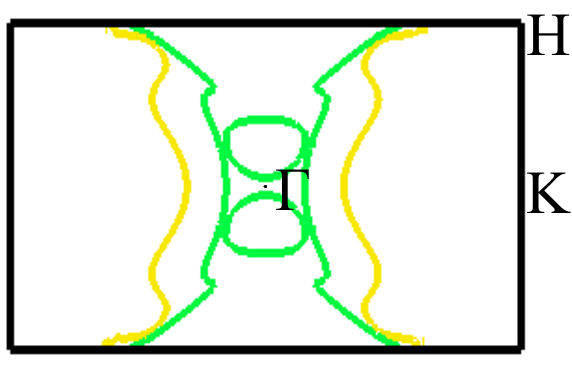}
    }
    \qquad
\subfloat[]{\label{fig:fzb}
\includegraphics[scale=0.45,trim=75px 0 75px 18px, clip]{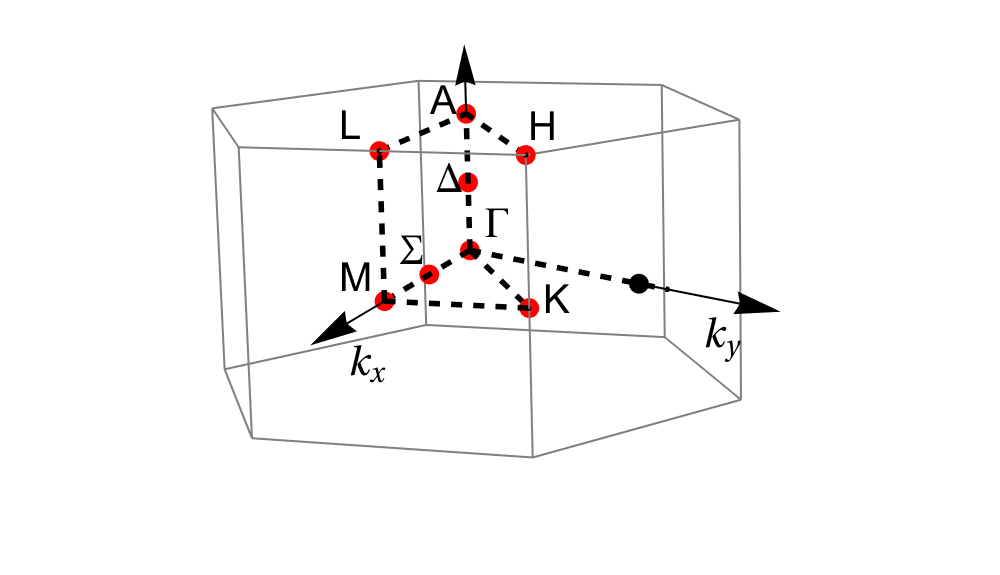}
}
\end{minipage}
 \caption{\label{fig:fermis}(a-d) Fermi surfaces obtained for the H$\parallel$z case. (a) There is a flat flower around $\Delta$-point while the ellipsoid around $\Gamma$ is not appreciated on this scale. (b) For this band, the FS is not simply connected and has two independent sheets, one showing a "nut"-shape in the extended zone, and the other one showing an "X"-shape centered in $K$-point, folded on itself and does not reach the top and bottom boundaries.
 (c) and (d) are FS having hyperboloid-like shape.
(e) Cross section of the band plotted in (b) in a extended plane that contains $\Gamma-K-M$ points, only closed orbits are allowed. (f) Cross section of the band plotted in (c)-yellow and (c)-green in a extended plane that contains $A-L-H$ points. The connections among yellow lines indicates that this surface holds open orbits in directions perpendicular to $c$-axis. (g) Cross section of the bands plotted in (b) and (c) in a  plane that contains $\Gamma-H-K$ points, it presents open orbits along the $c$-axis and a curvature that can be extrapolated to an hyperboloid-like surface cut.
(h) First Brillouin zone and its symmetric points used as reference.}
\end{figure*}

We found four bands crossing the Fermi level which lead to a FSs shape conformed of many sheets.
The calculated energy isosurface in the reciprocal space for each band that crosses the Fermi level in the  H$\parallel$y-case are presented in Fig. \subref{fig:fermi18}-\subref*{fig:fermi19}.
Our results indicate that the direction of the magnetic moments do not affect the shape of the FSs.
Nevertheless,  one situation has to be mentioned: the FS in Fig. \subref{fig:fermi18} has a small ellipsoid around $\Gamma$ and a ``flat flower" close to the $\Delta$ symmetric point.
This band is affected by the SOC in the H$\parallel$z-case, producing a small splitting arround $\Gamma$-point, leaving the Fermi level inside the gap.
As a consequence the ellipsoid sheet around $\Gamma$ disappears for the H$\parallel$z configuration.
The area of this ellipsoid is small compared with other FSs sheets and, consequently, its contribution to the MR and Hall conductivity at low temperatures is negligible.

The surface shown in Fig. \subref{fig:fermi21}  has two not-connected sheets, where the red part indicates the face of the surface with occupied states.
One of them has the shape of a "nut" when plotted in an extended BZ, but does not connect opposite borders, thus only closed orbits are allowed.
The other one has  an "X" shape made with a folded sheet on itself, but it does not reach the top and bottom borders.
Also, it has a negligible contribution to the conductivity tensor, since the normal vector  $d\mathbf{S}$ takes  opposite directions for close points in reciprocal space.
To visualize one possible orbit on this surface, we plot in Fig. \subref{fig:ext1} the cross-section in the repeated-zone of a plane perpendicular to $k_z$-axis that contains the $\Gamma$ point; it is clear that there is no possibility of holding open orbits.


\begin{figure*}[ht]
\begin{minipage}{0.2\textwidth}
  \subfloat[\label{fig:sample1}]{
   \includegraphics[keepaspectratio=true,scale=0.38]{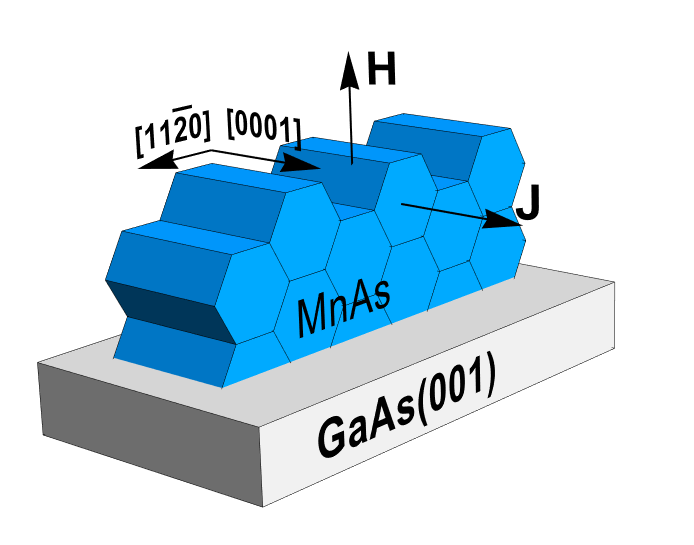}
  }\\
  \subfloat[\label{fig:sample2}]{
    \includegraphics[keepaspectratio=true,scale=0.38]{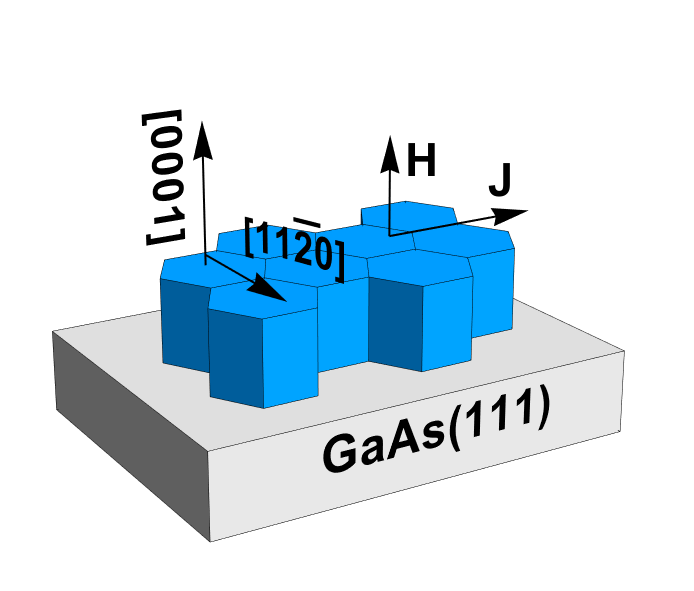}}
\end{minipage}
\begin{minipage}{0.75\textwidth}
 \subfloat[\label{fig:mr172}]{
    \includegraphics[keepaspectratio=true,scale=0.29]{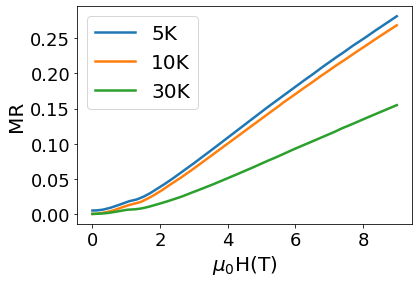}
 }
 \subfloat[\label{fig:hall172}]{
    \includegraphics[keepaspectratio=true,scale=0.29]{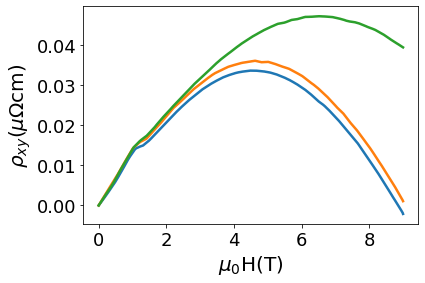}
 }
 \subfloat[\label{fig:m172}]{
    \includegraphics[keepaspectratio=true,scale=0.29]{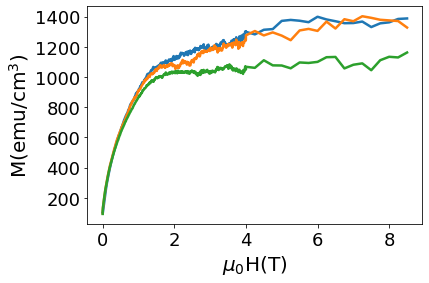}
 }
 \\
 \subfloat[\label{fig:mr171}]{
    \includegraphics[keepaspectratio=true,scale=0.29]{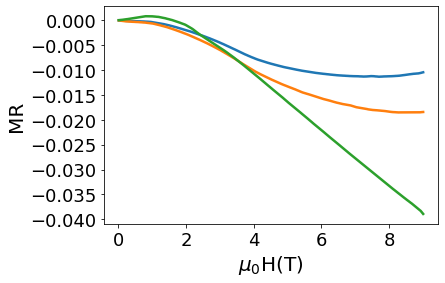}
 }
 \subfloat[\label{fig:hall171}]{
    \includegraphics[keepaspectratio=true,scale=0.29]{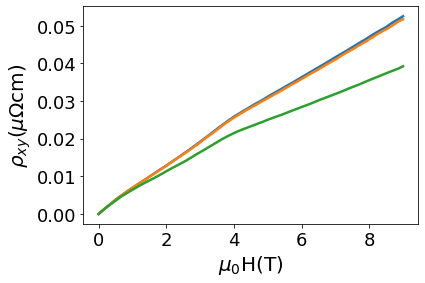}
 }
 \subfloat[\label{fig:m171}]{
    \includegraphics[keepaspectratio=true,scale=0.29]{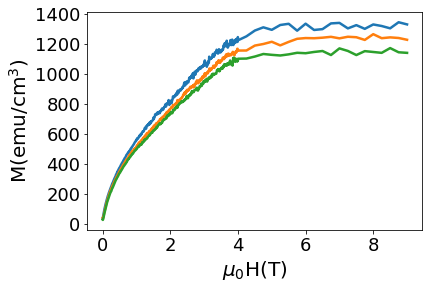}
 }
\end{minipage}
 \caption{\label{fig:transp} Magnetotransport measurements in MnAs/GaAs(001) (H$\parallel$y), upper row, and MnAs/GaAs(111) (H$\parallel$z), lower row, showing Magnetoresistance(c/f), Hall resistivity (d/g) and magnetization (e/h) for both epitaxies as a function of the applied magnetic field and temperature.
 The MR is positive and does not saturate in (c), while it is negative and saturates for low temperatures in (f).
 The Hall resistivity (d) shows a change in the carrier type manifested in the change of slope at $\sim$ 5-7 T, depending on T, that is not observed in (g), that shows a linear behavior. The shoulder observed at $\sim $ 1.5T in (d) and at $\sim$ 4T in (g)is related to the saturation of the magnetization in (e)and (h) at the corresponding fields, respectively.}
\end{figure*}

The FSs of interest are those shown in Fig. \subref{fig:fermi20} and \subref{fig:fermi19}.
Both surfaces are  hyperboloid-like along the $k_z$ -axis which is parallel to the $c$-crystal axis.
Now it becomes clear the association of these surfaces with the model presented in Sec. \ref{sec:teoria}: when the magnetization is parallel to the $c$-axis, it is related to the case where the magnetic field is parallel to the $z$-axis of the hyperboloid model (H$\parallel$z configuration).
On the other hand, when magnetization is parallel to the hexagonal planes, it relates to the H$\parallel$y configuration.
In Fig. \subref{fig:ext2} we present the cross section of the bands plotted in Figs. \subref{fig:fermi20} and \subref{fig:fermi19} at the repeated-zone in  a plane perpendicular to $k_z$-axis that contains the $A$-point. Interestingly, a path that contains open orbits in a direction perpendicular to the $k_z$-axis is observed.

The last two mentioned FSs have hyperboloid-like surface, which, as we describe in previous section, leads to a \textit{goniopolar} magnetotransport behavior that we corroborate with our experimental measurements.

\subsection{Magnetotransport measurements\label{sec:exp}}

The MnAs samples were epitaxially grown on GaAs(001) and GaAs(111) substrates \cite{Islam2017,Islam2018}, as depicted in the first column of Fig. \ref{fig:transp}.
These samples allow us to study the magnetotransport phenomena in two different MnAs-crystal orientations, where the applied magnetic field is parallel to the plane of the hexagon (H$\parallel$y) and the other one where the applied field is perpendicular to it (H$\parallel$z), as indicated in Fig. \ref{fig:transp}.
Magnetotransport data and magnetization measurements as a function of external magnetic field and temperature were done in a physical property measurement system (PPMS) using the Van der Pauw electric contact configuration \cite{Pauw1958}.

Results on the MnAs/GaAs(001) sample (H$\parallel$y) are displayed in the upper row of Fig. \ref{fig:transp}.
The MR presented in Fig. \subref{fig:mr172} is positive with positive slope for all temperatures, having a quasi-parabolic behavior, and does not saturate for the maximum applied field of 9 T.

The Hall resistivity  dependence with applied magnetic field \subref{fig:hall172} has a non monotonic  behaviour, showing a change in the slope around $\sim$ 5-7 T, that depends on the temperature. This behaviour has already been reported in Ref.\cite{Berry2001,Friedland2003} and was ascribed to a change on the type of carriers.
This behavior is the expected for FSs with hyperboloid shapes and magnetic field perpendicular to its symmetry axis as predicted in Sec. \ref{sec:teoria} and show in Fig. \ref{fig:rhoxy}.
The magnetization vs. field shown in Fig. \subref{fig:m172} rises until $\sim$ 1T and gradually saturates upon increasing the magnetic field.
The shoulder at $\sim$1 T in the Hall resistivity is related to the magnetization saturation at this field value, and indicates the saturation of the anomalous Hall effect \cite{Pugh1932}.

For the MnAs/GaAs(111) sample (H$\parallel$z-case), the MR presented in Fig.\subref{fig:mr171} has negative values and negative slope, and saturates for increasing applied magnetic field for 5 K and 10 K. Instead, for 30 K the saturation is not reached for the maximum applied field of 9 T.
This saturation observed in the low field regime indicates dominance of closed orbits in momentum space at low temperatures.
The Hall resistivity for MnAs/GaAs(111) sample displayed in Fig. \subref{fig:hall171} has a linear behaviour with increasing magnetic field, as expected for the hyperboloid-like FS with magnetic field parallel to symmetry axis.
There is a subtle change of slope at ~4 T, which is the same filed where the magnetisation saturates as shown in Fig. \subref{fig:m171}, suggesting the saturation of the anomalous Hall contribution to the Hall resistivity.

MBE growth ensures an excellent crystal quality as well as sharp interfaces between the MnAs layer and the GaAs substrate \cite{Daweritz2001}, allowing to rule out any significant role of the impurities at low temperatures. Also, possible magnetic domains disappear when magnetic saturation is reached.

To summarize the experimental results for different crystalline orientations, MnAs/GaAs(001) shows  a change of carrier type with increasing magnetic field, but for MnAs/GaAs(111) the carrier type is conserved. This means that the ordinary Hall behavior with magnetic field is in agreement with the model discussed in Sec. \ref{sec:teoria} for FSs like the ones presented in Sec.\ref{sec:electronic}. This is the main issue we addressed in this work.

\section{Discussion}
\begin{figure}[b]
 \includegraphics[scale=0.5]{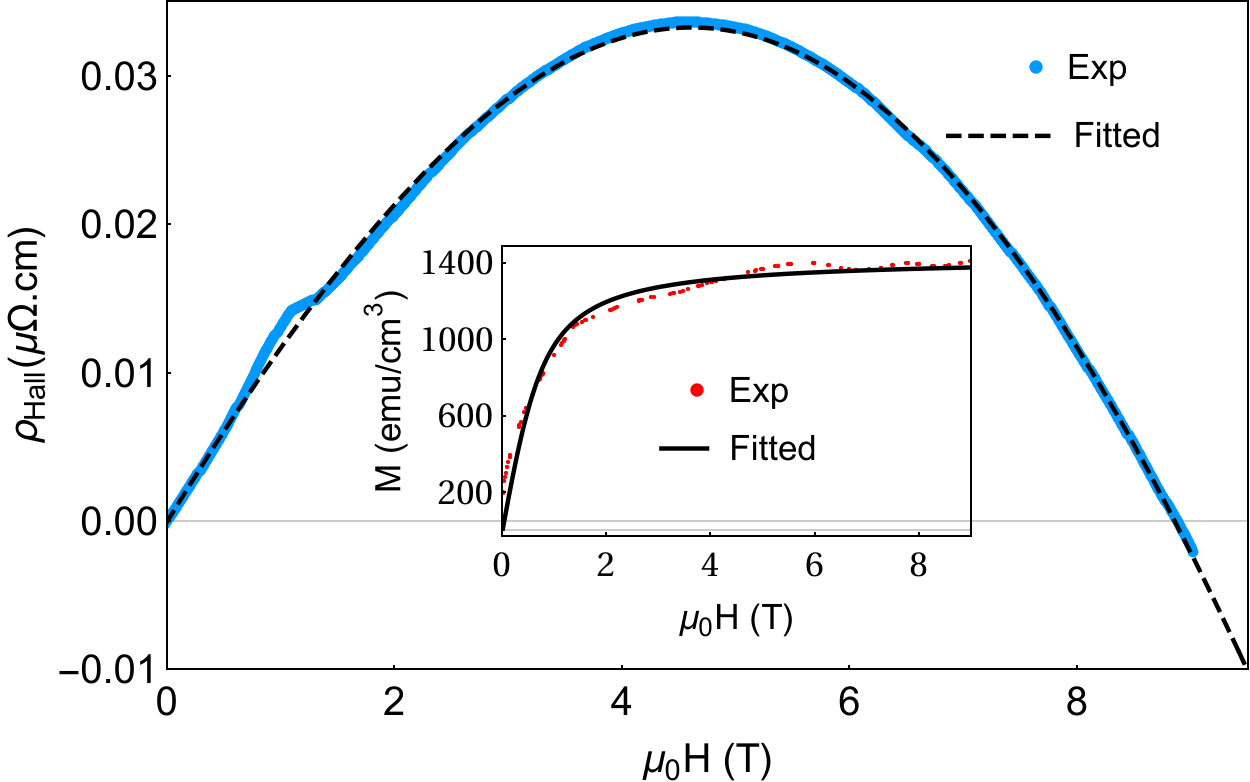}
 \caption{\label{fig:fit}Experimental measurement at 5 K of the Hall resistivity for sample MnAs/GaAs(001) fitted using Eq.\eqref{eq:rhotot}, where $\rho_{ord}=\rho_{xz}^{open}$.  At the inset the magnetization is fitted  by a Langevin function.}
\end{figure}

In order to reproduce the experimental results with the modeled FS, it is necessary take into account that the Hall resistivity measurements include two contributions, the ordinary Hall effect and the anomalous Hall effect. Our model considers only the first one.
They are related to the total Hall resistivity by \cite{Pugh1932,Chun2007}.
\begin{equation}
\label{eq:rhotot}
 \rho_{Hall}(H)=\rho_{ord}(H)+R_A\mu_0 M(H,T).
\end{equation}
Where $H$ and $M$ are the applied magnetic field and magnetization of the sample in the out of plane direction, respectively.
$\rho_{Hall}$ is the measured Hall resistivity, $R_A$ is the anomalous Hall coefficient and $\rho_{ord}$ is the ordinary Hall contribution, which is obtained from the non-diagonal elements of the Eq. \eqref{eq:abiertas} and \eqref{eq:cerradas}.

The Hall resistivity in MnAs/GaAs(111) presents a linear behavior with applied magnetic field (Fig. \subref{fig:hall171}).
This result is expected as we proved that in this configuration the possible orbits in reciprocal space are closed.
From Eq. \eqref{eq:cerradas} we obtain a linear behaviour for this case, where $\rho_{ord}=\rho_0.H$.
Instead, we predicted open orbits for the sample MnAs/GaAs(001), and our experimental results on Hall resistivity show a non linear behaviour with the applied magnetic field.
For this last case, we use $\rho_{x,z}^{open}$ from Eq. \eqref{eq:abiertas} as $\rho_{ord}$  in the Eq. \eqref{eq:rhotot} to reproduce the experimental results.

The agreement between the model and experimental Hall measurements for MnAs/GaAs (001) is shown in Fig.\ref{fig:fit}.
The fit yields a $g_2/\alpha\approx 2.10$ , which is comparable with the 4.09 and 1.41 values obtained for the FS of Figs. \subref{fig:fermi19} and \subref{fig:fermi20} respectively, parameterised using Eq.\eqref{eq:fs}. Finally, the coefficient $R_A=0.1\frac{n\Omega.cm}{T}$ has the same order of magnitude of similar metallic systems \cite{Yang2018}.

The results from the electronic structure calculation presented in Sec. \ref{sec:electronic} show a multiband contribution to the electron transport with four bands crossing the Fermi level.
However, one of the bands presented in Fig. \subref{fig:fermi18} has a negligible number of states compared with the other bands, which diminishes its contribution to conductivity.
Another band presented in Fig. \subref{fig:fermi21} is not a simply connected surface but it is made of two sheets that contribute with close orbits, although one of the sheets is folded on itself and its contribution to the conductivity is also negligible.
Finally, the remaining two bands have an hyperboloid-like shape along the $z$-axis in reciprocal space and they are the main contribution to the magnetorresistence dependence with magnetic field at low temperatures.

\section{Conclusion}

The experimental data evidence the different behaviour of the charge carriers depending on the crystal direction. For MnAs/GaAs(001) (H$\parallel$y configuration) the carrier polarity changes when the system goes from low field state ($\omega\tau<1$) to high field state ($\omega\tau>1$), while this change of carrier polarity is not observed in MnAs/GaAs(111) (H$\parallel$z configuration).
We modeled this behaviour using the shape of the FS in MnAs.
The specific topology of the dominant FSs sheets with an hyperboloid shape and the good agreement between measurements and the model allow us to classify this material as \textit{goniopolar}; and also to understand the ordinary Hall behavior in this compound with magnetic field, that remained elusive for more than a decade.
We believe our study provides guidelines to study the magnetotransport properties in a broad range of materials with similar FSs.

\section{Acknowledgments}

We acknowledge D. P. Daroca, C. Gourdon, L. Thevenard, D. Hrabovsky and Y. Klein for the valuable discussions and their assistance in the magneto-transport measurements at the "Plateforme de mesures physiques a basse temperature" at Sorbonne Universite.
Samples were grown by E. Islam  and M. Akabori at Center for Nano Materials and Technology (CNMT), Japan Advanced Institute of Science and Technology (JAIST), 1-1 Asahidai, Nomi, Ishikawa 923-1292,Japan. This work was supported by CONICET (PIP 11220115-0100213 CO), ANPCyT (PICT-2016-0867) and LIFAN (Laboratoire International Franco Argentin en Nanosciences) collaboration.

\bibliographystyle{unsrtnat}
\bibliography{Goniopolar2021.bib}

\begin{thebibliography}{27}
\providecommand{\natexlab}[1]{#1}
\providecommand{\url}[1]{\texttt{#1}}
\expandafter\ifx\csname urlstyle\endcsname\relax
  \providecommand{\doi}[1]{doi: #1}\else
  \providecommand{\doi}{doi: \begingroup \urlstyle{rm}\Url}\fi

\bibitem[Pippard(1960)]{Pippard1960}
A.~B. Pippard.
\newblock {Experimental analysis of the electronic structure of metals}.
\newblock \emph{Rep. Prog. Phys.}, 23\penalty0 (1):\penalty0 176--266, 1960.
\newblock ISSN 00344885.
\newblock \doi{10.1088/0034-4885/23/1/304}.

\bibitem[{M. Liftshitz, M.I. Azbel}(1956)]{1955}
M.I.~Kaganov {M. Liftshitz, M.I. Azbel}.
\newblock {On the theory of Galvanomagnetic Effects in metals}.
\newblock \emph{JETP}, 3\penalty0 (1):\penalty0 143--146, 1956.

\bibitem[Gurzhi and Kopeliovich(1975)]{Gurzhi1975}
R~N Gurzhi and A~I Kopeliovich.
\newblock {Galvanomagnetic properties of metals with closed Fermi surfaces at
  low temperatures}.
\newblock \emph{JETP}, 40:\penalty0 1144--1151, 1975.

\bibitem[Gurzhi and Kopeliovich(1977)]{Gurzhi1977}
N~Gurzhi and I~Kopeliovich.
\newblock {Galvanomagnetic properties of metals with open Fermi surfaces}.
\newblock 44\penalty0 (2):\penalty0 334--340, 1977.

\bibitem[Novikov et~al.(2019)Novikov, {De Leo}, Dynnikov, and
  Maltsev]{Novikov2019}
S.~P. Novikov, R.~{De Leo}, I.~A. Dynnikov, and A.~Ya Maltsev.
\newblock {Theory of Dynamical Systems and Transport Phenomena in Normal
  Metals}.
\newblock \emph{J. Exp. Theor. Phys.}, 129\penalty0 (4):\penalty0 710--721,
  2019.
\newblock ISSN 10906509.
\newblock \doi{10.1134/S106377611910008X}.

\bibitem[He et~al.(2019)He, Wang, Arguilla, Cultrara, Scudder, Goldberger,
  Windl, and Heremans]{He2019}
Bin He, Yaxian Wang, Maxx~Q. Arguilla, Nicholas~D. Cultrara, Michael~R.
  Scudder, Joshua~E. Goldberger, Wolfgang Windl, and Joseph~P. Heremans.
\newblock {The Fermi surface geometrical origin of axis-dependent conduction
  polarity in layered materials}.
\newblock \emph{Nat. Mater.}, 2019.
\newblock ISSN 1476-1122.
\newblock \doi{10.1038/s41563-019-0309-4}.
\newblock URL \url{http://www.nature.com/articles/s41563-019-0309-4}.

\bibitem[Wang et~al.(2020)Wang, Koster, Ochs, Scudder, Heremans, Windl, and
  Goldberger]{Wang2020}
Yaxian Wang, Karl~G. Koster, Andrew~M. Ochs, Michael~R. Scudder, Joseph~P.
  Heremans, Wolfgang Windl, and Joshua~E. Goldberger.
\newblock {The Chemical Design Principles for Axis-Dependent Conduction
  Polarity}.
\newblock \emph{J. Am. Chem. Soc.}, 142\penalty0 (6):\penalty0 2812--2822,
  2020.
\newblock ISSN 15205126.
\newblock \doi{10.1021/jacs.9b10626}.

\bibitem[Radha and Lambrecht(2020)]{Lambrecht2020}
Santosh~Kumar Radha and Walter R.~L. Lambrecht.
\newblock {Topological band structure transitions and goniopolar transport in
  honeycomb antimonene as a function of buckling}.
\newblock \emph{Phys. Rev. B}, 101:\penalty0 235111, Jun 2020.
\newblock \doi{10.1103/PhysRevB.101.235111}.
\newblock URL \url{https://link.aps.org/doi/10.1103/PhysRevB.101.235111}.

\bibitem[Zhang et~al.(2019)Zhang, Wu, Liu, and Yazyev]{Zhang2019}
Shengnan Zhang, Quansheng Wu, Yi~Liu, and Oleg~V. Yazyev.
\newblock {Magnetoresistance from Fermi surface topology}.
\newblock \emph{Phys. Rev. B}, 99\penalty0 (3):\penalty0 1--12, 2019.
\newblock ISSN 24699969.
\newblock \doi{10.1103/PhysRevB.99.035142}.

\bibitem[Berry et~al.(2001)Berry, Potashnik, Chun, Ku, Schiffer, and
  Samarth]{Berry2001}
J.~J. Berry, S.~J. Potashnik, S.~H. Chun, K.~C. Ku, P.~Schiffer, and
  N.~Samarth.
\newblock {Two-carrier transport in epitaxially grown MnAs}.
\newblock \emph{Phys. Rev. B - Condens. Matter Mater. Phys.}, 64\penalty0
  (5):\penalty0 2--5, 2001.
\newblock ISSN 1550235X.
\newblock \doi{10.1103/PhysRevB.64.052408}.

\bibitem[Friedland et~al.(2003)Friedland, K{\"a}stner, and
  D{\"a}weritz]{Friedland2003}
K.~J. Friedland, M.~K{\"a}stner, and L.~D{\"a}weritz.
\newblock {Ordinary Hall effect in MBE-grown MnAs films grown on GaAs(001) and
  GaAs(111)B}.
\newblock \emph{Phys. Rev. B - Condens. Matter Mater. Phys.}, 67\penalty0
  (11):\penalty0 4, 2003.
\newblock ISSN 1550235X.
\newblock \doi{10.1103/PhysRevB.67.113301}.

\bibitem[Berger(1969)]{Berger1969}
L.~Berger.
\newblock {Hall effect of a compensated magnetic metal proportional to MB2 in
  the high-field limit}.
\newblock \emph{Phys. Rev.}, 177\penalty0 (2):\penalty0 790--792, 1969.
\newblock ISSN 0031899X.
\newblock \doi{10.1103/PhysRev.177.790}.

\bibitem[D{\"a}weritz et~al.(2001)D{\"a}weritz, Schippan, Trampert,
  K{\"a}stner, Behme, Wang, Moreno, Sch{\"u}tzend{\"u}be, and
  Ploog]{Daweritz2001}
L~D{\"a}weritz, F~Schippan, A~Trampert, M~K{\"a}stner, G~Behme, Z.M Wang,
  M~Moreno, P~Sch{\"u}tzend{\"u}be, and K.H Ploog.
\newblock {MBE growth, structure and magnetic properties of MnAs on GaAs on a
  microscopic scale}.
\newblock \emph{Journal of Crystal Growth}, 227-228:\penalty0 834--838, 2001.
\newblock ISSN 0022-0248.
\newblock \doi{10.1016/S0022-0248(01)00897-1}.
\newblock URL \url{http://www.sciencedirect.com/science/article/pii/
  S0022024801008971}.
\newblock Proceeding of the Eleventh International Conference on Molecular Beam
  Epitaxy.

\bibitem[Ramsteiner et~al.(2002)Ramsteiner, Hao, Kawaharazuka, Zhu,
  K{\"a}stner, Hey, D{\"a}weritz, Grahn, and Ploog]{Ramsteiner2002}
M.~Ramsteiner, H.~Y. Hao, A.~Kawaharazuka, H.~J. Zhu, M.~K{\"a}stner, R.~Hey,
  L.~D{\"a}weritz, H.~T. Grahn, and K.~H. Ploog.
\newblock {Electrical spin injection from ferromagnetic MnAs metal layers into
  GaAs}.
\newblock \emph{Phys. Rev. B}, 66:\penalty0 081304, Aug 2002.
\newblock \doi{10.1103/PhysRevB.66.081304}.
\newblock URL \url{https://link.aps.org/doi/10.1103/PhysRevB.66.081304}.

\bibitem[Mattoso et~al.(2004)Mattoso, Eddrief, Varalda, Ouerghi, Demaille,
  Etgens, and Garreau]{Mattoso2004}
N.~Mattoso, M.~Eddrief, J.~Varalda, A.~Ouerghi, D.~Demaille, V.~H. Etgens, and
  Y.~Garreau.
\newblock {Enhancement of critical temperature and phases coexistence mediated
  by strain in MnAs epilayers grown on $\mathrm{Ga}\mathrm{As}(111)B$}.
\newblock \emph{Phys. Rev. B}, 70:\penalty0 115324, Sep 2004.
\newblock \doi{10.1103/PhysRevB.70.115324}.
\newblock URL \url{https://link.aps.org/doi/10.1103/PhysRevB.70.115324}.

\bibitem[Ashcroft and Mermin(1976)]{Ashcroft1976}
Neil~W. Ashcroft and N.~David Mermin.
\newblock \emph{{Solid state physics}}.
\newblock Holt-Saunders, international ed edition, 1976.

\bibitem[Akgoz and Saunders(1975)]{Akgoz1975}
Y.~C. Akgoz and G.~A. Saunders.
\newblock {Space-time symmetry restrictions on the form of transport tensors.
  II. Thermomagnetic effects}.
\newblock \emph{J. Phys. C Solid State Phys.}, 8\penalty0 (18):\penalty0
  2962--2970, 1975.
\newblock ISSN 00223719.
\newblock \doi{10.1088/0022-3719/8/18/016}.

\bibitem[Tanaka(2002)]{Tanaka2002}
Masaaki Tanaka.
\newblock {Ferromagnet ({MnAs})/{III}~V semiconductor hybrid structures}.
\newblock \emph{Semiconductor Science and Technology}, 17\penalty0
  (4):\penalty0 327--341, mar 2002.
\newblock \doi{10.1088/0268-1242/17/4/306}.
\newblock URL \url{https://doi.org/10.1088%2F0268-1242%2F17%2F4%2F306}.

\bibitem[Tanaka et~al.(1994)Tanaka, Harbison, Park, Park, Shin, and
  Rothberg]{Tanaka1994}
M.~Tanaka, J.~P. Harbison, M.~C. Park, Y.~S. Park, T.~Shin, and G.~M. Rothberg.
\newblock {Epitaxial orientation and magnetic properties of MnAs thin films
  grown on (001) GaAs: Template effects}.
\newblock \emph{Applied Physics Letters}, 65\penalty0 (15):\penalty0
  1964--1966, 1994.
\newblock \doi{10.1063/1.112831}.

\bibitem[Rungger and Sanvito(2006)]{Rungger2006}
Ivan Rungger and Stefano Sanvito.
\newblock {Ab initio study of the magnetostructural properties of MnAs}.
\newblock \emph{Phys. Rev. B - Condens. Matter Mater. Phys.}, 74\penalty0
  (2):\penalty0 1--14, 2006.
\newblock ISSN 10980121.
\newblock \doi{10.1103/PhysRevB.74.024429}.

\bibitem[Giannozzi et~al.(2017)]{Giannozzi2017}
P~Giannozzi et~al.
\newblock {Advanced capabilities for materials modelling with Quantum
  {ESPRESSO}}.
\newblock \emph{Journal of Physics: Condensed Matter}, 29\penalty0
  (46):\penalty0 465901, oct 2017.
\newblock \doi{10.1088/1361-648x/aa8f79}.
\newblock URL \url{https://doi.org/10.1088%2F1361-648x%2Faa8f79}.

\bibitem[Islam and Akabori(2017)]{Islam2017}
Md.~Earul Islam and Masashi Akabori.
\newblock {Growth and magnetic properties of MnAs/InAs hybrid structure on
  GaAs(111)B}.
\newblock \emph{Journal of Crystal Growth}, 463:\penalty0 86--89, 2017.
\newblock ISSN 0022-0248.
\newblock \doi{10.1016/j.jcrysgro.2017.02.009}.
\newblock URL \url{http://www.sciencedirect.com/science/article/pii/
  S0022024817300830}.

\bibitem[Islam and Akabori(2018)]{Islam2018}
Md.~Earul Islam and Masashi Akabori.
\newblock {In-plane isotropic magnetic and electrical properties of
  MnAs/InAs/GaAs(111)B hybrid structure}.
\newblock \emph{Physica B: Condensed Matter}, 532:\penalty0 95--98, 2018.
\newblock ISSN 0921-4526.
\newblock \doi{10.1016/j.physb.2017.03.013}.
\newblock URL \url{http://www.sciencedirect.com/science/article/pii/
  S0921452617301175}.
\newblock Special issue on Frontiers in Materials Science: Condensed Matters.

\bibitem[van~der Pauw(1958)]{Pauw1958}
L.~J. van~der Pauw.
\newblock {A method of measuring specific resistivity and Hall effect of discs
  of arbitrary shape}.
\newblock \emph{Phylips Res. Repts}, 13:\penalty0 1--9, February 1958.

\bibitem[Pugh and Lippert(1932)]{Pugh1932}
E.~M. Pugh and T.~W. Lippert.
\newblock {Hall e.m.f. and Intensity of Magnetization}.
\newblock \emph{Phys. Rev.}, 42:\penalty0 709--713, Dec 1932.
\newblock \doi{10.1103/PhysRev.42.709}.
\newblock URL \url{https://link.aps.org/doi/10.1103/PhysRev.42.709}.

\bibitem[Chun et~al.(2007)Chun, Kim, Choi, Jeong, Lee, Suh, Oh, Kim, Khim, Woo,
  and Park]{Chun2007}
S.~H. Chun, Y.~S. Kim, H.~K. Choi, I.~T. Jeong, W.~O. Lee, K.~S. Suh, Y.~S. Oh,
  K.~H. Kim, Z.~G. Khim, J.~C. Woo, and Y.~D. Park.
\newblock {Interplay between carrier and impurity concentrations in annealed
  Ga1-xMnxAs: Intrinsic anomalous hall effect}.
\newblock \emph{Phys. Rev. Lett.}, 98\penalty0 (2):\penalty0 1--4, 2007.
\newblock ISSN 10797114.
\newblock \doi{10.1103/PhysRevLett.98.026601}.

\bibitem[Yang et~al.(2018)Yang, Luo, Wu, Xu, Li, Pennycook, Zhang, and
  Wu]{Yang2018}
Yumeng Yang, Ziyan Luo, Haijun Wu, Yanjun Xu, Run~Wei Li, Stephen~J. Pennycook,
  Shufeng Zhang, and Yihong Wu.
\newblock {Anomalous Hall magnetoresistance in a ferromagnet}.
\newblock \emph{Nat. Commun.}, 9\penalty0 (1):\penalty0 1--9, 2018.
\newblock ISSN 20411723.
\newblock \doi{10.1038/s41467-018-04712-9}.
\newblock URL \url{http://dx.doi.org/10.1038/s41467-018-04712-9}.

\end{thebibliography}
\end{document}